\documentstyle[12pt,worldsci]{article}
\newcommand{\avg}[1]{\langle #1 \rangle}

\def\twi#1{\widetilde{#1}}

\def\Re{{ \mbox{Re} \,}}
\def\Im{{ \mbox{Im} \,}}
\def\qw{Q_\ssw}
\def\ol#1{\overline{#1}}
\def\Pl{\gamma_\lft}
\def\lft{{\sss L}}
\def\rht{{\sss R}}
\def\Pr{\gamma_\rht}
\def\tw#1{\tilde{#1}}
\def\twVL{\tw{V}_\ell}
\def\twVQ{\tw{V}_q}

\def\etal{{\it et.al.}}
\def\ie{{\it i.e.}}

\def\nn{\nonumber}
\def\roughlyup#1{\mathrel{\raise.3ex\hbox{$\sim$\kern-.75em
\lower1ex\hbox{$#1$}}}}
\def\roughlydown#1{\mathrel{\raise.3ex\hbox{$#1$\kern-.75em
\lower1ex\hbox{$\sim$}}}}

\def\lsim{\roughlydown<}
\def\gsim{\roughlydown>}
\def\simeq{\roughlyup-}

\def\sss{\scriptscriptstyle}

\def\cc{{\rm cc}}

\def\mw{M_{\sss W}}
\def\mz{M_{\sss Z}}
\def\gf{G_{\sss F}}

\def\rht{{\sss R}}
\def\lft{{\sss L}}

\def\twg{\tw{g}}
\def\twh{\tw{h}}

\def\sw{s_w}
\def\cw{c_w}

\def\dpi{\delta \Pi}

\def\ssv{{\sss V}}
\def\ssa{{\sss A}}
\def\ssl{{\sss L}}
\def\ssw{{\sss W}}

\def\ssh{{\sss H}}
\def\ssz{{\sss Z}}
\def\mw{m_{\sss W}}
\def\mz{m_{\sss Z}}
\def\gf{G_{\sss F}}
\def\rht{{\sss R}}
\def\lft{{\sss L}}
\def\sm{{\sss SM}}

\def\ww{{\sss WW}}
\def\aa{{\gamma\gamma}}
\def\zz{{\sss ZZ}}
\def\zg{{{\sss Z}\gamma}}
\def\sw{s_w}
\def\cw{c_w}

\def\dgz{\Delta g_{1\ssz}}

\def\dkz{\Delta \kappa_\ssz}
\def\dkg{\Delta \kappa_\gamma}

\def\lz{\lambda_\ssz}
\def\lg{\lambda_\gamma}

\def\eqa{\begin{eqnarray}}
\def\eeqa{\end{eqnarray}}
\def\eq{\begin{equation}}
\def\eeq{\end{equation}}

\def\Scl{{\cal L}}

\def\nth#1{{1 \over #1}}

\begin{document}

\rightline{NEIP-94-012, McGill-94/50}
\rightline{hep-ph/9411257}

%\rightline{September 1994}

\title{{\bf THE EFFECTIVE USE OF PRECISION ELECTROWEAK MEASUREMENTS}}
\author{C.P. BURGESS\thanks{Invited talk presented to the 3rd Workshop
on High-Energy Particle Physics, Madras India, January 1994.} \\
\vspace{0.3cm}
{\em Institut de Physique, Universit\'e de Neuch\^atel \\
1 Rue A.L. Breguet, CH-2000 Neuch\^atel, Switzerland.}\\
\vspace{0.2cm}
and \\
\vspace{0.2cm}
{\em
Physics Department, McGill University \\
3600 University St.,  Montr\'eal, Qu\'ebec, Canada, H3A 2T8.}\\
}
\maketitle
\setlength{\baselineskip}{2.6ex}

{\begin{center} ABSTRACT \end{center}
{\small \hspace*{0.3cm}
Several reasonably model-independent formulations of the implications
of new physics for precision electroweak measurements have been developed
over the past years, most notably by Peskin and Takeuchi, and by Altarelli
\etal.
These formulations work by identifying a small, but useful, set of
parameters through which new physics often enters into well-measured physical
observables. For the theories to which such an analysis applies, this approach
greatly streamlines the confrontation with the data. Since the
experimentally-allowed range for these parameters has been determined
from global fits to the data, theorists need
only compute their predictions for these parameters to constrain their models.
We summarize these methods here, together with several recent generalizations
which permit applications to wider classes of new physics, and which include
the original approaches as special cases. }}

\section{Introduction}

The attainment of high precision in measurements of $Z$-boson properties
has been perhaps the most significant experimental result in high-energy
physics over the last decade. Besides testing the Standard Model (SM) to high
precision, experiments at LEP and at SLC \cite{LEP} are providing the first
experimental winnowing of the bumper crop of theories that hope to describe the
physics at energies well above 100 GeV.

Indeed, the data on the $Z$ resonance is now so good that model builders
ignore it at their peril. There is a very real benefit in confronting
the various models of current theoretical interest with the constantly
improving experimental results. One way to proceed is to simply compute
the relevant observables explicitly on a model-by-model basis, and to fit
the results to the data in order to constrain the model's parameter space.
Unfortunately, this is a time-consuming procedure and so one is limited
in the number of models which can be treated in this way. For well-motivated
theories, such as the minimal supersymmetric generalizations to the standard
model  (MSSM) for instance, such detailed calculations may be worth the
effort they require,  although even here it is impractical to explore the
model's
entire parameter space.

Happily, there is another way to proceed which can substantially reduce the
labour that is required to confront a model with the implications of the data.
This alternative is based on the realization that many models often only
contribute
to deviations from the SM --- for the well-measured observables of interest ---
in a limited number of ways. For instance, this could happen if all of the new
particles only couple to the presently-observed ones in a restricted
manner, such as through the exchange of
electroweak gauge bosons. Or all of the new particles could be extremely heavy.
In either case it is typically true that only a small number of independent
combinations of the model's coupling constants ever appear in --- and so are
well constrained by --- the observables that are measured. In these cases it
is useful to confront the data in two steps. First, one can parameterize the
well-measured observables in terms of a few independent variables which can
then be constrained, once and for all, by comparing with the experiments. Next,
constraints on any given model may be obtained by comparing the bounds on
these parameters with the model's predictions for them as functions of
its underlying couplings.

Such a two-step procedure has the advantage of separating the statistical fit
to
the data from model-dependent calculations. Since the data can be fit, once and
for all, to a general set of parameters, it is not necessary to repeat this
analysis
separately for every model. Although, in principle, the permitted parameter
space that is obtained for a particular theory in this way can differ somewhat
from what would be obtained by a direct fit to the model, in practice the two
procedures turn out to give constraints which are essentially equivalent.
Furthermore, the comparative model-independence of the fit to the data
in the two-step approach permits an efficient comparison of many models,
and so gives a reasonably broad picture of the kinds of new physics which
can produce deviations in different observables.

There are a number of similar, but not completely equivalent, examples
of this type of reasoning which have become widely used in the
literature. \citea{
\citenum{PT},\citenum{MR},\citenum{KLang},\citenum{AB},\citenum{HMHK}}
The most widely used of these are
based on the parameterizations of Peskin \etal\ \cite{PT}, and
of Altarelli \etal\ \cite{AB}. Both of these formalisms are very useful for
describing the implications for precision electroweak measurements of a
wide class of new physics. Neither of these parameterizations of the
observables
can encompass all models, however, and so it is important to bear in mind
the limits to their applicability when attempting to use their results for
comparing with any particular theory. Some well-motivated models and
observables cannot be analysed completely within the framework of either
approach, motivating their extension to more general
situations. \citea{\citenum{alphabet},\citenum{bigfit}} The purpose of this
review is to describe those extensions which I have helped to develop, and
to relate these extensions to earlier approaches. Some applications of the
resulting techniques are also described.

This article is organized as follows. There are two main categories of new
physics that have been considered to date: those which dominantly contribute to
observables through `oblique' corrections, and those for which the new physics
is heavy. The next two sections are devoted to describing how each of these
kinds of corrections can contribute to electroweak observables. In the next
section the case of `oblique' corrections is considered, while heavy,
non-oblique, new physics is considered in section 3. Section 4 briefly
describes
a number of the applications of these techniques, and section 5 summarizes the
conclusions.

\section{Oblique Physics}

Almost all of the observables that are presently amenable to accurate
measurement can be phrased in terms of the two-particle scattering of
light fermions. This is because these experiments either involve the scattering
of two quarks or leptons, or the decay of an initial fermion into three lighter
ones. There are three ways in which new physics can affect such experiments,
given that the new particles are not themselves directly produced. It can:
($a$) change the propagation of the gauge bosons that can be exchanged by the
fermions; ($b$) alter the three-point fermion -- boson couplings; and ($c$)
modify the four-point direct fermion -- fermion interactions (\ie:
`box'-diagram
corrections).

An important class of new-physics models contribute dominantly to precision
measurements through process ($a$): changes to the vacuum polarizations of the
electroweak bosons. Such corrections are called `oblique' \cite{LPS}, and when
they dominate they imply {\it universal} modifications to light-particle
scattering, in the sense that the changes depend only on the electroweak
quantum numbers of the light
fermion involved.  Oblique corrections can be the most important when the
direct
couplings between the observed light fermions and any new particles are either
forbidden or highly suppressed.

\subsection{General Oblique Corrections}

The effects of oblique corrections on fermion scattering can be determined by
examining how the gauge boson vacuum polarizations
\eq
\Pi^{\mu\nu}_{ab}(q) = \Pi_{ab}(q^2) \; \eta^{\mu\nu} + (q^\mu q^\nu
\mbox{  terms}),
\eeq
(with $a,b = \gamma,W,Z$) appear in the observables of interest.
\citea{\citenum{LPS},\citenum{KL}} The contribution to these due to
new physics we denote as
$\dpi_{ab}(q^2)$, so the full vacuum polarization is given by: $\Pi_{ab}(q^2) =
\Pi^\sm_{ab}(q^2) + \dpi_{ab}(q^2)$.

In general, each of the $\dpi_{ab}(q^2)$ can be arbitrary functions of $q^2$,
and so the vacuum polarizations in principle contain several unknown functions,
each of which can potentially enter into all physical observables. Any attempt
to extract general information by fitting these functions to the data might
therefore seem to be doomed because of the large number of unknown quantities
in comparison to the amount of data that is available. This turns out to be too
pessimistic a view, however, because currently accurate measurements are
performed either at very low energies, or on the $Z$ resonance: \ie\ only for
$q^2 = \mz^2$ and $q^2 \approx 0$. ($q^2 = \mw^2$ should also be considered to
the extent that the mass and width of the $W$ boson are thought to be
sufficiently well measured.) As a result, the unknown functions,
$\dpi_{ab}(q^2)$, are presently only accurately sampled at these few values of
four-momentum transfer.

This restriction --- that precision observables only probe $q^2 \approx 0$ and
$q^2
= \mz^2$ and $\mw^2$ --- implies that all oblique corrections to electroweak
observables can be expressed in terms of six independent combinations of the
various $\dpi$'s \cite{alphabet}. The counting proceeds as follows.

\begin{enumerate}
\item %1.
Inspection of the graphs with vacuum polarization insertions shows that, {\it a
priori}, there are ten quantities to consider. For neutral-current processes at
$q^2=0$ and $q^2 = \mz^2$ these are: $\dpi_\aa(q^2)/q^2$, $\dpi_\zg(q^2)/q^2$
and $\dpi_\zz(q^2)$. (Notice that electromagnetic gauge invariance ensures that
both $\dpi_\aa(q^2)/q^2$ and $\dpi_\zg(q^2)/q^2$
are well-defined at $q^2=0$.) $\dpi'_\zz(\mz^2)$, with the prime denoting
differentiation with respect to $q^2$, also contributes at $q^2 = \mz^2$.
Finally, charged-current observables, or those involving physical $W$
particles,
involve  $\dpi_\ww(q^2)$ at $q^2=0$ and $\mw^2$, as well as $\dpi'_\ww(\mw^2)$.
\item % 2.
Of these ten possible parameters, three combinations can never lead to
observable deviations from the SM, since they can be absorbed into SM
renormalizations. For instance, they can be absorbed into renormalizations for
the electroweak gauge potentials, $W^a_\mu$, $B_\mu$, and of the Higgs {\it
vev}, $\avg{\phi}$. This reduces the total number of possible oblique
parameters to seven.
\item % 3.
Finally, at the present levels of accuracy, new-physics contributions to
$\dpi_{\gamma\gamma} (\mz^2)$ are not detectable. This is because this term
contributes purely through photon exchange, which is not resonantly enhanced
when $q^2 = \mz^2$. As a result, the influence of $\dpi_{\gamma\gamma}
(\mz^2)$ is suppressed by $O\left( \Gamma_{\ssz} / \mz \right) \sim 0.03$
in comparison to the effects of the $Z$-mediated terms $\dpi_{\ssz\gamma}
(\mz^2)$ and $\dpi_{\ssz\ssz} (\mz^2)$. We are therefore left with six
measurable oblique parameters.
\end{enumerate}

There is obviously a great deal of freedom in how to parameterize
this six-dimensional parameter space. A convenient way to define
the six oblique parameters is:
\eqa
{\alpha S \over 4 \sw^2 \cw^2 } &=& \left[\, {\dpi_\zz(\mz^2) - \dpi_\zz(0)
\over \mz^2} \,\right] - {(\cw^2 - \sw^2) \over \sw\cw} \;
\delta{\widehat\Pi}_\zg(0) - \delta{\widehat\Pi}_\aa(0), \nn\\
\alpha T &=& {\dpi_\ww(0) \over \mw^2} - {\dpi_\zz(0) \over \mz^2}, \nn\\
{\alpha U \over 4 \sw^2 } &=& \left[\, {\dpi_\ww(\mw^2) - \dpi_\ww(0) \over
\mw^2} \, \right] - \cw^2 \left[ \, { \dpi_\zz(\mz^2) - \dpi_\zz(0) \over
\mz^2} \, \right] \nn \\
&&\qquad \qquad  -  \sw^2 \delta{\widehat\Pi}_\aa(0) - 2 \sw \cw
\delta{\widehat\Pi}_\zg(0) \\
\alpha V &=& \dpi_\zz'(\mz^2) - \left[ \, {\dpi_\zz(\mz^2) - \dpi_\zz(0)
\over \mz^2}  \, \right], \nn\\
\alpha W &=& \dpi_\ww'(\mw^2) - \left[ \, {\dpi_\ww(\mw^2) - \dpi_\ww(0)
\over \mw^2}  \, \right],  \nn\\
\alpha X &=& - \sw \cw \left[ \, \delta{\widehat\Pi}_\zg(\mz^2) -
\delta{\widehat\Pi}_\zg(0) \, \right], \nn
\eeqa
where $\sw$ and $\cw$ denote the sin and cosine of the weak mixing angle,
$\theta_w$, and $\alpha$ represents the electromagnetic fine-structure
constant. For numerical purposes we later use $\alpha(\mz^2)=1/128$ and
$\sw^2=0.23$. The quantity $\delta{\widehat\Pi}_{ab}(q^2)$ denotes the ratio
$\delta\Pi_{ab}(q^2)/q^2$. These definitions are chosen so that ($i$) the
first three agree with the definitions of $S$, $T$ and $U$ that are used
in Ref.~\citenum{MR}, ($ii$) the remaining three quantities, $V$,
$W$ and $X$, vanish if $\dpi_{ab}(q^2)$ should be simply a linear
function of $q^2$, and ($iii$) each parameter contributes to a particular
kind of observable (see below).

A straightforward calculation gives expressions for the corrections to the
various electroweak observables in terms of the parameters $S$ through $X$.
One complication arises because these corrections must be referred to the
corresponding (radiatively-corrected) SM prediction. Because of the necessity
for performing radiative corrections, the expressions for the SM predictions,
in
turn, depend on ($i$) which three observables are used to infer the
experimental
values for the SM couplings, and ($ii$) unknown quantities, such as the masses,
$m_t$, $m_\ssh$,  of the $t$ quark and Higgs boson.

We follow the universal practice of using the three best-measured
observables --- $\alpha$ from low-energy electron properties,
$\gf$ from muon decay, and $\mz$ from LEP --- as inputs for fixing
the values of the SM couplings from experiment. How the parameters,
$S$ through $X$, appear in the final expressions depends in detail upon
this choice, since new physics also affects these input observables, and so
shifts the inferred values for the SM couplings. With these inputs, and
with the definitions of $S$ through $X$ given above, $U$ and $W$ only
enter into `charged-current' quantities like the mass and width of the
$W$ boson, and of these $W$ appears only in the $W$-boson width.
Purely neutral-current data depend only on $S$, $T$, $V$
and $X$, and of these only $S$ and $T$ appear in observables at low
energies (for which $q^2 \approx 0$). $V$ and $X$ arise only in observables
defined at the $Z$ resonance. Of these two, $X$ enters as a correction to
the effective weak mixing angle, and so contributes to asymmetries such
as $A_{\sss LR}$ or $A_{\sss FB}$. $V$, on the other hand, drops out
of these asymmetries, but instead changes the $Z$ partial widths, since
it alters the normalization of the $Z$-fermion couplings. The numerical
expressions for these observables as functions of $S$ through $X$
are summarized in Table I.

The expressions from Table I can now be compared with the data to obtain bounds
on the phenomenologically allowed range for the parameters $S$ through $X$.
In making this comparison we use the values $m_t = 150$ GeV and $m_\ssh =
300$ GeV for computing the SM prediction. The sensitivity of our results to
these assumptions is addressed in section 4. The one-$\sigma$ allowed ranges
for the oblique parameters which result from the fit to the data of
Ref.~\citenum{alphabet} are then given by:
\eqa
&& S = -0.93 \pm 1.7; \qquad T = -0.67 \pm 0.92 ; \qquad U = -0.6 \pm 1.1;
\nn\\
&& V = 0.47 \pm 1.0 ; \qquad W = 1.2 \pm 7.0 ; \qquad X = 0.10\pm 0.58.
\eeqa
Notice that since the parameter $W$ only appears in the width of the
$W$ boson, it is the most poorly constrained.

\subsection{When the New Physics is Heavy}

There is an important special case for which the above oblique analysis
simplifies considerably. This is when the lightest mass, $M$, for all of the
new
particles is much larger than $\mz$. Since the scale over which $q^2$ varies
appreciably in $\dpi_{ab}(q^2)$ is set by $M$, in this case these functions may
be well approximated by the first terms in their Taylor expansion in powers of
$q^2/M^2$. Since current measurements are restricted to the regime $q^2 \lsim
\mz^2$, this approximation is controlled by powers of the small parameter
$\mz^2/M^2$.

The leading contributions in this limit --- \ie\ those which are not suppressed
by inverse powers of $M^2$ --- are simply linear functions
\citea{\citenum{PT},\citenum{MR},\citenum{KLang}}
of $q^2$:
\eq
\dpi_{ab}(q^2) \approx A_{ab} + B_{ab} \, q^2 .
\eeq
(Higher-order terms have also been considered in the
literature. \cite{GW}) With this
assumption three of the oblique parameters, $V$, $W$ and $X$, vanish
identically and so all new physics effects are described by the three
parameters
$S$, $T$ and $U$. The expressions for observables in terms of these three
parameters may therefore be obtained simply by setting $V = W = X = 0$ in Table
I.  A fit to the data, \cite{alphabet}  with $V, W$ and $X$ constrained to
vanish, then gives the following one-$\sigma$ allowed ranges for $S$, $T$ and
$U$:
\eq
S = -0.48\pm 0.40 ; \qquad T = -0.32\pm 0.40 ; \qquad U = -0.12\pm 0.69.
\eeq
Not surprisingly, the allowed range for $S$ and $T$ in this two-parameter fit
is smaller than was permitted in the six-parameter fit whose results are given
above.

Since the neutral-current data are completely controlled by the two parameters
$S$ and $T$, it has become conventional to display the results of fits to this
data by plotting the ellipses of constant confidence interval in the
two-dimensional $S$-$T$ plane. Figure 1 displays the results of the two fits
described above. Notice that, since $\mw$ can depend on the parameter $U$
as well as on $S$ and $T$, $\mw$ should only be included in such a plot
if there are {\it a priori} reasons for believing $U$ to be negligibly small.

\subsection{Using Only the $Z$ Resonance}

In recent years the electroweak data on the $Z$ resonance has become more
accurate than are the older low-energy measurements. As a result it is now
possible \cite{AB} to usefully constrain new physics using {\it only} the data
at $q^2 = \mz^2$. As is clear from the earlier counting of parameters, such a
restriction to only one value of four-momentum transfer permits a description
of
the data in terms of fewer oblique parameters than the six that were required
when $q^2 \approx 0$ was also considered. This is true even if the new
particles
associated with the new physics are not heavy compared to $\mz$.

As is easily verified by repeating the counting argument given earlier, only
three parameters are required to describe the general oblique corrections
in this case. A convenient choice for these three parameters is: \cite{negs}
\eqa
S' &=& S + 4 \sw^2 \cw^2 \; V + 4 (\cw^2 - \sw^2) \; X , \nn\\
T' &=& T + V ,\\
U' &=& U - 4 \sw^2 \cw^2 V + 8 \sw^2 X . \nn
\eeqa
With this choice, the parameters $S'$, $T'$ and $U'$ reduce to $S$, $T$ and $U$
in the limit where $V = W = X = 0$. This ensures that the dependence of all
$Z$-pole observables on $S'$, $T'$ and $U'$ can be simply read off from Table
I by replacing $(S,T,U,V,W,X) \to (S',T',U',0,0,0)$.

As before, since the data on the $Z$ resonance only depends on two
parameters $S'$ and $T'$, it is convenient to display the result of a fit of
these two parameters to the data as a plot in the $S'$-$T'$ plane. The result
of such a fit to the $Z$ data only \cite{steve} is displayed
in Figure 2.

It is noteworthy that the constraints on $S'$ and $T'$ from Figure 2 are
comparable to those on $S$ and $T$ using the larger data set, including
also the $q^2 \approx 0$ observables, that are used for Figure
1. This illustrates the high quality of the data that has been obtained over
recent years at the $Z$ resonance. The conceptual difference between Figures 1
and 2 remains crucial, however. Whereas a two-parameter, $S'-T'$, description
of the $Z$-pole data relies only on the assumption that oblique corrections
dominate, the extension of such a description to include also
neutral-current measurements at $q^2 \approx 0$ relies on the additional
assumption that all of the new physics is sufficiently massive to permit the
neglect of $V$ and $X$.

\subsection{A Cultural Aside}

Treating the $Z$-pole data by itself is very much in the spirit of the approach
of
Ref.~\citenum{AB}. The formalism of these authors differs from that
described so far in the following two ways, however.

\begin{enumerate}
\item %1
These authors introduce three parameters, $\epsilon_1$, $\epsilon_2$, and
$\epsilon_3$, which broadly correspond to the three parameters $S'$, $T'$ and
$U'$. Their definitions, however, are made directly in terms of the
observables,
and do not separate the new-physics contributions from those due to SM
radiative corrections. As a result only the
deviation, $\delta \epsilon_i = \epsilon_i - \epsilon_\sm$, from the SM
predictions --- using the fiducial choices for $m_t$ and $m_\ssh$ --- are
directly related to $S'$, $T'$ and $U'$. The connection between the two sets of
parameters is: \citea{\citenum{AB},\citenum{nevans},\citenum{ivan}}
\eq
\delta \epsilon_1 = \alpha T', \qquad  \delta\epsilon_2 = - \; { \alpha U'
\over 4 \sw^2},
\qquad  \delta \epsilon_3 = {\alpha S' \over 4 \sw^2}.
\eeq
\item %2
A second important difference in the approach of Altarelli \etal\ is to permit
one non-oblique correction, parameterized by $\epsilon_b$, to the $Zbb$ vertex.
The inclusion of this correction is motivated by the large $m_t$-dependent
contributions it receives from SM radiative corrections, and potentially from
other kinds of new physics. We return to this type of term in the next
section.
\end{enumerate}

\section{Nonoblique New Physics}

Although many kinds of new physics dominantly produce oblique corrections to
electroweak observables, this is certainly not true for all kinds. Examples
include any new physics which preferentially couples to, say, the heavy
generations. Another worthwhile generalization of the previous formalism is
therefore to extend it to include nonoblique corrections. In order to be
practical, however, such a generalization cannot be permitted to introduce too
many new parameters, or else the utility of the confrontation with the data
will be lost. Some criterion is necessary to limit and organize the number of
independent interactions that need be considered.

A very natural way to provide the required organization is to assume that
all of the new physics is much heavier than the electroweak scale, $\mz$.  In
this case the implication of such new physics for current experiments can be
parameterized in terms of a low-energy effective lagrangian \cite{elreviews}
such as would be obtained by integrating out all of the presently-undiscovered
heavy particles. In this section the results of such an analysis \cite{bigfit}
are summarized.

The limit of large masses, $M$, for all hypothetical new particles allows their
low-energy interactions to be organized according to dimension, with operators
having a  higher mass dimension being more suppressed by inverse powers of $M$.
Simple dimension counting need not be the whole story, of course, as low-energy
selection rules and symmetries can also help to determine the relative size of
the various effective interactions. For a more complete discussion of the
issues
involved see Ref.~\citenum{bigfit}.

\subsection{The Effective Interactions}

Consider, therefore, the most general effective interactions that are
consistent
with the particle and symmetry content that is appropriate for applications to
processes having energies $\lsim 100$ GeV. Since our intention is to study
current experiments in this energy range, we take our particle content to
include only those which already have been detected. This includes most of
the SM particles, including precisely three left-handed neutrinos, but does
{\it
not} include the Higgs boson and the top quark, which we take to have been
integrated out (if they indeed exist \cite{top}). Due to the absence of these
particles, the electroweak gauge group must be nonlinearly realized on the
given fields, and so in practice it can be completely ignored \cite{ignored}
(apart from the unbroken electromagnetic subgroup) in what follows. The price
for choosing this particle content is that the resulting effective lagrangian
has to violate unitarity at energies at or below the TeV range.

Typically, any such lagrangian contains a great many effective interactions.
Fortunately, there is a great deal of latitude in how the interactions can be
written, since there is considerable freedom to redefine fields to simplify
terms in the lagrangian. We choose to work with fields for which the kinetic
and
mass terms take their standard diagonal forms. The
nonstandard interactions which arise in the most general effective
lagrangian up to mass dimension five can then be written in the following
way.\cite{bigfit}

\bigskip\noindent
{\it 1. Fermion Masses:}
The only possible effective interactions having dimension two or three are
an arbitrary set of masses for the $W$ and $Z$ bosons and for each of the
low-energy fermions. The fermion mass terms so obtained are indistinguishable
from those which appear in the SM, with the exception of any neutrino masses. A
neutrino mass matrix would have two effects. It would ($i$) give the neutrino
mass eigenstates nonzero masses, and ($ii$) it would introduce unitary
Cabbibo-Kobayashi-Maskawa
(CKM) style mixing matrices, $\twVL^{ij}$ into the various charged-current
neutrino couplings. (Flavour off-diagonal neutrino kinetic terms, such as
can arise when sterile neutrinos are integrated out, can also
introduce off-diagonal neutral-current interactions, and can make the
charged-current mixing matrices nonunitary.)
Notice that, unlike most other new-physics corrections,
these mixing angles need not be small, even if the neutrino masses are.

\bigskip\noindent
{\it 2. Electromagnetic Couplings:}
The total electromagnetic couplings of fermions are straightforward to
write down:
\eq
\Scl_{\rm em} = -e  \Bigl[ \ol{f}_i \gamma^\mu Q_i \, f_i \; A_\mu +
  \ol{f}_i \sigma^{\mu\nu} ( d^{ij}_\lft \Pl + d^{ij}_\rht \Pr)
  \, f_j \;  F_{\mu\nu} \Bigr],
\eeq
where the indices $i$ and $j$ are to be summed over all possible flavours of
light fermions, $f_i$. $Q_i$ represents the electric charge of $f_i$, in units
of the proton charge. $\Pl$ and $\Pr$ denote the usual projection matrices
onto left- and right-handed spinors. Linear combinations of the effective
coupling matrices, $d^{ij}_\lft$ and $d^{ij}_\rht$, represent nonstandard
magnetic- and electric-dipole moment interactions.

\bigskip\noindent
{\it 3. Charged-Current Interactions:}
The fermion charged-current interactions become:
\eqa
\Scl_{\rm cc} &=&  -{e\over \sqrt{2}\sw} \; \Bigl[ \ol{f}_i \gamma^\mu
    (h_\lft^{ij} \Pl + h_\rht^{ij} \Pr )\, f_j \; W^*_\mu  \nn\\
    && \qquad \qquad + \ol{f}_i \sigma^{\mu\nu}
    ( c_\lft^{ij} \Pl + c_\rht^{ij}
    \Pr) \, f_j \; W^*_{\mu\nu} \Bigr] + \cc,
\eeqa
where $W_{\mu\nu} = D_\mu W_\nu - D_\nu W_\mu$ is the $W$ field strength
using electromagnetic covariant derivatives, $D_\mu$.  For leptons,
\eqa
h_\lft^{\nu_i \ell_j} &=&  \delta \twh_\lft^{\nu_i \ell_j}  \nn \\
&& + \twVL^{ij} \left(1 - { \alpha S \over 4
( \cw^2 - \sw^2)} + { \cw^2\; \alpha T \over 2 (\cw^2 - \sw^2)}  + {\alpha
U \over 8 \sw^2} - { \cw^2\; (\Delta_e + \Delta_\mu) \over 2 (\cw^2 -
\sw^2)} \right), \nn\\
h_\rht^{\nu_i \ell_j} &=& \delta \twh_\rht^{\nu_i \ell_j}~,
\eeqa
while for quarks:
\eqa
h_\lft^{u_i d_j} &=& \delta \twh_\lft^{u_i d_j} \nn \\
&&  + \twVQ^{ij} \left(1 - { \alpha S \over 4
   ( \cw^2 - \sw^2)} + { \cw^2\; \alpha T \over 2 (\cw^2 - \sw^2)}  +
{\alpha U \over 8 \sw^2} - { \cw^2\; (\Delta_e + \Delta_\mu) \over 2 (\cw^2
- \sw^2)} \right), \nn\\
h_\rht^{u_i d_j} &=& \delta \twh_\rht^{u_i d_j}.
\eeqa
Here $\twVL^{ij}$ and $\twVQ^{ij}$ respectively denote the unitary
CKM matrices for
the left-handed charged current interactions of the leptons and quarks. The
coefficients $\delta \twh_{\sss L(R)}^{u_i d_j}$ and $c^{ij}_{\sss L(R)}$
represent a set of arbitrary nonstandard fermion-$W$ couplings, and $S$, $T$
and
$U$ are defined in terms of the vacuum polarizations as in the previous
section.
Finally, $\Delta_f$ (with $f=e$, $\mu$ or $\tau$) denotes the following
quantity:
$\Delta_f \equiv \sqrt{\sum_i \left| h_\lft^{\nu_i f}  \right|^2} \; - 1 $.

\bigskip\noindent
{\it 4. The $W$ Mass:}
Although the SM contains gauge boson mass terms, these arise only in a
particular linear combination, leading to a calculable mass relation for $\mw$
in terms of the three inputs, $\mz$, $\gf$ and $\alpha$: $\mw =
\mw^\sm \equiv \mz \cw + $(radiative corrections). This relation is
ruined by the generic effective gauge-boson mass terms, leading to the
following nonstandard contribution to the $W$ mass: \cite{bigfit}
\eq
\mw^2 = (\mw^\sm)^2 \left[ 1 - {\alpha S \over 2 ( \cw^2 - \sw^2)}  +
   {\cw^2 \; \alpha T \over  \cw^2 - \sw^2}  + { \alpha U \over 4 \sw^2}
   - {\sw^2 (\Delta_e + \Delta_\mu) \over \cw^2 - \sw^2} \right] .
\eeq

\bigskip\noindent
{\it 5. Neutral-Current Couplings:}
The general interactions between the light fermions and the $Z$ boson can be
written as follows:
\eq
\Scl_{\rm nc} = -{e\over \sw \cw} \; \Bigl[ \ol{f}_i \gamma^\mu (
   g_\lft^{ij} \Pl + g_\rht^{ij} \Pr ) \, f_j
   \; Z_\mu + \ol{f}_i \sigma^{\mu\nu} ( n_\lft^{ij} \Pl + n_\rht^{ij}
   \Pr) \, f_j  \; Z_{\mu\nu} \Bigr],
\eeq
where
\eqa
 \Bigl( g_\lft^{ij} \Bigr)_\sm &=&  \Bigl( T_{3i} - Q_i \,
   \sw^2 \Bigr) \; \delta^{ij} \nn\\
  \Bigl( g_\rht^{ij} \Bigr)_\sm &=&  \Bigl( - Q_i \, \sw^2 \Bigr)
    \; \delta^{ij} \nn\\
 g_{\sss L(R)}^{ij} &=& \Bigl( g_{\lft,\rht}^{ij} \Bigr)_\sm \;
   \left[ 1 + {1 \over 2} ( \alpha T - \Delta_e  - \Delta_\mu )
   \right] \\
   && - Q_i \, \delta^{ij}
   \; \left({ \alpha S \over 4 ( \cw^2 - \sw^2)} - { \cw^2\sw^2 \; \alpha T
    \over \cw^2 - \sw^2} + {\cw^2 \sw^2 (\Delta_e + \Delta_\mu) \over
   \cw^2 - \sw^2} \right) \nn \\
  && + \delta \twg_{\sss L(R)}^{ij} .
\eeqa
As before, $Q_i$ represents here the electric charge of fermion $f_i$, and
$T_{3i}$ is its eigenvalue for the third component of weak isospin.
$Z_{\mu\nu}$
denotes the abelian curl: $\partial_\mu Z_\nu - \partial_\nu Z_\mu$. The
effective coupling matrices $\delta \twg_{\sss L(R)}^{ij}$ and $n_{\sss
L(R)}^{ij}$
represent arbitrary sets of nonstandard couplings between the fermions and the
$Z$ boson.

These expressions may now be used to compute electroweak observables. Working
to linear order in the effective couplings permits the neglect of all
interactions which cannot interfere with the corresponding SM contribution.
This eliminates a good many of the effective vertices that are listed
above, including most flavour-changing interactions.
(Ref.~\citenum{bigfit} gives a more
general discussion which also includes the strongly bounded
flavour-changing couplings.) The results for a number of low-energy
observables (those for which $q^2 \approx 0$) are listed in Table II, and
those for observables at the weak scale in Table III.

The parameters which appear in Tables II and III can be fit to the precision
electroweak data, just as was done for the oblique parameters $S$ through $X$.
The results of such a fit \cite{bigfit} are quoted in Tables IV, V and VI. In
these
tables the results of two types of fits are presented. In one of these (the
`Individual
Fit')  the parameter in question has been considered in isolation,
with all of the other parameters set to zero by hand. This kind of fit is not
realistic, but has often been considered in the literature. The second fit
(the `Global Fit') allows all of the parameters to be varied in fitting the
data.
Perhaps surprisingly, the resulting bounds on the various parameters are
nevertheless quite good.

\subsection{The $Zbb$ Vertex}

An important special case of the above analysis is to simply add a nonstandard
dimension-four coupling between the $Z$ boson and the $b$ quark, in addition
to the oblique corrections $S$, $T$ and $U$. This corresponds to the approach
of
Ref.~\citenum{AB}, with the parameters $\epsilon_1, \epsilon_2$ and
$\epsilon_3$ given as in the previous section, and $\epsilon_b$ given by:
\eq
\delta \epsilon_b \equiv \epsilon_b - \epsilon_b^\sm = -2 \;  \delta
\twg_\lft^{bb}.
\eeq
Notice that, like all of the flavour-diagonal neutral-current couplings,
the reality of the lagrangian implies that $\delta \twg_\lft^{bb}$ must
be real. In this case, the formulae of Tables II and III simplify considerably.

\section{Applications}

With the above results in place for the experimental limits on the various
effective parameters, the next step is to compute the values of these
parameters
in terms of the couplings and masses of various underlying models. We therefore
now turn to a brief summary of some of the applications to which the above
formalism has been made.

\subsection{$m_t$ Dependence}

The starting point for the comparison with experiment is the choice of a set of
fiducial values, $\hat m_t$ and $\hat m_\ssh$, for the unknown masses,
$m_t$ and $m_\ssh$, which appear in the SM contributions to various
observables.
We have chosen $\hat m_t \equiv 150$ GeV and $\hat m_\ssh \equiv 300$ GeV in
performing the fits, but it is natural to wonder how the SM predictions
change as these masses vary.

Perhaps the simplest, and most useful, application of the techniques described
in earlier sections is to approximately determine this $m_t$ and $m_\ssh$
dependence. To do  so, imagine both the top quark and the Higgs boson to be
much
heavier than $\mz$. In this case, all of the implications for lower-energy
observables due to these heavy particles can be phrased in terms of the
effective lagrangian obtained by integrating them out of the SM. It can
therefore be mimicked by choosing an appropriate $m_t$ and $m_\ssh$ dependence
for some of the effective couplings of section 3.

Evaluating the one-loop vacuum polarization graphs \citea{\citenum{PT},
\citenum{MR},\citenum{KLang},\citenum{AB}} containing
top and Higgs loops gives the following large-mass dependence for the oblique
parameters:
\eqa
\delta S_\sm &\simeq& - \; {1 \over 3 \pi} \; \ln \left( { m_t \over
\hat m_t} \right) + {1 \over 6 \pi} \;  \ln \left({ m_\ssh \over \hat m_\ssh}
\right) \nn \\
\delta T_\sm &\simeq& {3 \over 16 \pi \sw^2 \cw^2 } \; \left( {m_t^2 - \hat
m_t^2
\over \mz^2} \right) + {3 \over 8 \pi \cw^2} \; \left[ \ln
\left({m_t \over \hat m_t} \right)  - \ln \left( { m_\ssh \over
\hat m_\ssh} \right) \right] \nn \\
\delta U_\sm &\simeq& - \; {1 \over \pi} \;  \ln \left( { m_t \over \hat m_t}
\right) .
\eeqa
Similarly, the $Zbb$ vertex-correction graph involving a virtual top quark
gives \citea{\citenum{AB},\citenum{topcorrections}}
\eq
\Bigl( \delta \twg^{bb}_\lft \Bigr)_\sm \simeq {\alpha \over 8 \pi \sw^2 \cw^2}
\; \left( {m_t^2 - \hat m_t^2 \over \mz^2} \right)  .
\eeq
These are the {\it only} virtual contributions which can both ($i$) depend
strongly, at one loop, on $m_t$, and ($ii$) contribute appreciably to
electroweak
observables.

\subsection{Exotic Fermions}

One application to which the above formalism has been applied \cite{bigfit} is
the  determination of the constraints on the masses and couplings that are
possible for hypothetical exotic fermions which can mix with the ordinary
ones. For the present purposes we take `exotic' to mean new fermions which
transform under the $SU_\ssl(2)$ gauge group as either left-handed singlets
and/or right-handed doublets. Because of their mixing with ordinary fermions,
exotic fermions of this sort can change the couplings of the ordinary fermions
to the $W$ and $Z$. Constraints on these mixings can be inferred from present
precision measurements of these couplings.

One motivation for doing this analysis, is that it also has been performed
\cite{exotic} by fitting these models directly to the data. By comparing the
bounds obtained here with those of the direct fit to the model we can learn
how closely they agree with one another. One would expect the
present approach to give bounds that are marginally weaker than those of a
direct fit to the model, since there are typically more parameters in the
general effective lagrangian than there are couplings in a particular
underlying theory. In this sense our results can be considered the most
conservative bounds possible. The interesting point is that the limits we
obtain
are not very much weaker at all, and so very little information is lost by
using
the simpler effective-lagrangian fit. For simplicity of presentation we
restrict
ourselves here to charged quarks, although neutrino mixing can be handled in
much the same way \cite{bigfit}.

In general, mixing between the known fermions and exotic ones induces
flavour-changing processes into the neutral current interactions of
ordinary fermions (FCNC's). For simplicity we ignore these types of induced
couplings here, although they can be treated in a similar manner. FCNC's are
avoided if every known electroweak multiplet of fermions mixes separately with
its own exotic partner, in which case the mixing can be characterized in terms
of a mixing angle, $\theta^i_{\lft(\rht)}$, $i=e,\mu,\tau,u,d,s,\dots$.

To make contact with our general formalism, simply integrate out all of the
undiscovered exotic particles to produce the low-energy effective theory.
Sufficient accuracy is obtained by working at tree level, and so it is easy
to integrate out the heavy fermions: one transforms to a basis of mass
eigenstates, and sets all heavy fields equal to zero. As a result we find
$S$=$T$=$U$=0, and the nonstandard neutral current and charged-current
couplings
of the ordinary fermions become modified by \cite{bigfit}:
\eq
\delta \twg_\lft^{ii} = - T_{3i} \left(s_\lft^i\right)^2
{}~,\qquad \delta \twg_\rht^{ii} = + T_{3i} \left(s_\rht^i\right)^2 ~,
\eeq
and
\eq
\delta \twh_\lft^{u_i d_j} = -{1\over 2} \, \twVQ^{ij}
\left[ (s_\lft^{u_i})^2 + (s_\lft^{d_j})^2 \right]~, \qquad
\delta \twh_\rht^{u_i d_j} = s_\rht^{u_i} s_\rht^{d_j}
\twi{U}_\rht^{ij}~,
\eeq
in which $\left( s^i_{\lft(\rht)} \right)^2 \equiv 1 - \left( c^i_{\lft(\rht)}
\right)^2 \equiv \sin^2 \theta_{\lft(\rht)}^i$, and $\twi{U}_\rht^{ij}$ is
a unitary CKM-type matrix for the right-handed charged-current couplings.
Similar expressions also hold for leptons \cite{bigfit}.

It is now a simple matter to bound the mixing angles using these expressions
together with the constraints of Tables V and VI.
We find the following limits at 90\% c.l.~(defined as 1.64$\sigma$):
\eqa
\Delta_{e,\mu} : && \left(s_\lft^e\right)^2, \left(s_\lft^{\nu_e}\right)^2 <
0.016 ;
\qquad  \left(s_\lft^\mu\right)^2, \left(s_\lft^{\nu_\mu}\right)^2 < 0.012
\nn\\
\delta \twh_\lft^{ud} : && \left(s_\lft^u\right)^2, \left(s_\lft^d\right)^2 <
0.02 \\
\delta g_{\lft,\rht}^{ii} : && \left(s_\rht^e\right)^2 < 0.01; \quad
 \left(s_\rht^\mu\right)^2 < 0.09; \quad \left(s_\rht^u\right)^2 < 0.03 ; \nn
\\
&&\left(s_\rht^d\right)^2 < 0.05 ;
\quad \left(s_\lft^s\right)^2 < 0.05 ; \quad
\left(s_\lft^b\right)^2 < 0.03 , \nn
\eeqa
A comparison of the above numbers with those in the literature \cite{exotic}
confirms that they are very similar to, but marginally weaker than, those found
by fitting directly to the mixing angles themselves.

\subsection{Light Exotic Particles}

Another application that has been examined with this formalism
\citea{\citenum{negs},\citenum{nevans},\citenum{hfam}} is
the case of new exotic particles whose quantum numbers preclude their mixing
with ordinary fermions \cite{extragauge}. Such particles arise in a great
many types of theories for new physics, including supersymmetric models,
theories with additional generations, and technicolour models.
In these examples the new particles need {\it not} be much heavier
than the weak scale, and so we do not make this assumption here.

Unlike for the previous example, in this case the absence of tree-level mixing
implies that the dominant effects of the new physics arise through its one-loop
contributions to well-measured observables. This is in most cases dominated by
the contributions to gauge boson vacuum polarizations, and so we consider here
only oblique corrections.

Formulae for the one-loop contributions to the six oblique parameters, $S - X$,
by scalars and fermions in general representations of the electroweak gauge
group have been given in the literature
\citea{\citenum{alphabet},\citenum{scalars},\citenum{techniscalars}}.
For example, the one-loop vacuum polarization due to a fermion with left- and
right-handed coupling constants, $k^a_{\lft(\rht)}$, to gauge boson `$a$' (with
$a = \gamma,W,Z$) is:
\eq
\dpi_{ab}(q^2) = {1 \over 2 \pi^2} \sum_{ij} \int_0^1 dx \; f_{ab}(q^2,x)
\; \ln \left[ { m^2_{ij}(x) - q^2 x (1-x) \over \mu^2 } \right] . \eeq
Here $m^2_{ij}(x) \equiv m^2_i (1-x) + m^2_j x$, where $m_i$ and $m_j$ are
the masses of the two fermions which appear within the loop, and
\eqa
f_{ab}(q^2,x) &=& {k^a_{\lft} k_{\lft}^{b*} + k_{\rht}^a k_{\rht}^{b*} \over 2}
\left[ x (1-x) \, q^2 - { m^2_{ij}(x) \over 2} \right] \nn\\
&& \qquad + {k_{\lft}^a k_{\rht}^{b*} + k_{\rht}^a k_{\lft}^{b*} \over 2}
\; \left( { m_i m_j \over 2} \right).
\eeqa
For example, a standard-model doublet would have $k_{\lft}^\gamma =
k_{\rht}^\gamma = e Q_i$, $k_{\lft}^\ssz = (e/\sw\cw) [T_{3i} - Q_i \sw^2]$,
$k_{\rht}^\ssz = (e/\sw\cw) [ - Q_i \sw^2]$,  $k_{\lft}^\ssw = e / \sqrt{2}
\sw$ and
$k_{\rht}^\ssw = 0$. $\mu^2$ denotes the renormalization point, where we have
renormalized using dimensional regularization and $\ol{\hbox{MS}}$.

Expressions such as these permit a survey of the couplings and masses that are
permitted for exotic particles transforming in a variety of electroweak gauge
representations \cite{negs}. For the purposes of illustration, Figure 3
displays the range of values for $S'$ and $T'$ that are produced by an
additional generation of quarks and leptons. Notice that this region does {\it
not} necessarily include the origin $(S' = T' = 0)$ in the limit that the
members
of this additional generation become very heavy. This is because heavy fermions
like these need not decouple from $S'$ and $T'$ in this limit. This makes
the oblique parameters especially sensitive probes for these types of new
particles.

\subsection{Is Technicolour Dead?}

The above observation, that some heavy particles need not decouple from
oblique parameters like $S$ and $T$ as their masses get very large, has been
used to argue that technicolour-like models for dynamical electroweak symmetry
breaking are ruled out by the current electroweak data
\citea{\citenum{PT},\citenum{MR},\citenum{KLang},\citenum{AB}}.
Various estimates for the oblique parameters in these theories predict
$S' \gsim O(+1)$, and the two-parameter  $S - T$ fit to the $Z$-pole data
excludes these values at roughly the three-$\sigma$ level. (A similar
conclusion has been reached by considering the anomalous $Zbb$
coupling as well \cite{technib}.)

This conclusion would be inescapable provided that these parameters
could be reliably computed to be $O(+1)$ in technicolour models.
Unfortunately, since these are strongly coupled theories, the robustness
of the various estimates is not clear \cite{negatives}.
For example, a potential loophole may exist when light particles are
present in the technispectrum --- such as often happens in these theories
due to the appearance of pseudogoldtone bosons. In this case the analysis
requires the additional parameters $V$ through $X$, and the
contributions of the light particles {\it can} be made
to contribute negatively to $S'$ and $T'$
\citea{\citenum{techniscalars},\citenum{negs}}, for some choices
for their masses and mixings.

These loopholes are difficult to completely close due to the difficulty
in computing with these theories. In the meantime, it must be conceded that
the case against these theories remains persuasive, albeit circumstantial.

\subsection{TGV's}

As a final application we consider how the oblique analysis of section 2 can be
used to bound \cite{tgvbounds} the potential existence of nonstandard
self-couplings for the electroweak gauge bosons (TGV's) \cite{HPZH}.
Somewhat surprisingly, the full six-parameter set of oblique corrections turns
out to be required for this analysis, even though all heavy particles are
assumed to be much more massive than the weak scale. The need for all six
parameters follows because loop-induced effects to oblique parameters are at
most of order $\alpha/4 \pi$, and this is the same order of magnitude as is
$\mz^2/M^2$. As a result, TGV loop-induced corrections to the $STU$ parameters
can be the same size as the other quantities, $VWX$, and so these must be
properly included.

The starting assumption is that the lightest mass, $M$, of any undiscovered
particle is high in comparison with the weak scale, $\mz$. Integrating out
this heavy physics gives an effective lagrangian, defined at the scale $M$,
which we assume to include the five CP-conserving TGV's, whose coefficients
are, in a standard \cite{HPZH} notation: $\dkg$, $\dkz$, $\dgz$, $\lz$, and
$\lg$. The goal is to constrain the coefficients of these effective
interactions
using precision electroweak measurements at lower energies, $E \lsim \mz$.

In order to do so we imagine running the effective lagrangian defined at the
scale $M$, down to the scale $\mz$ where current measurements are made. Since
anomalous TGV's are not directly probed at these energies, the bounds come
from the other effective interactions which are generated by the TGV's during
the running down from $M$ to $\mz$. We therefore compute all of the effective
interactions at the weak scale that get generated (at one loop) by loops
containing TGV's.

There are two kinds of effective interactions that are produced by TGV's
in this way: ($i$) oblique parameters, and ($ii$) nonstandard
fermion/gauge-boson
vertices. The fermion/gauge-boson vertices themselves come in two types: ($a$)
those which are universal, in that they are independent of the masses of the SM
fermions, and ($b$) those which depend strongly on the top-quark mass, $m_t$.
Interestingly, we find that all of the vertex corrections of type ($a$) that
are generated by TGV's can be rewritten as oblique corrections by performing a
suitable field redefinition \cite{tgvbounds}. The entire analysis can
therefore be done using the oblique parameters, together with a small number
of $m_t$-dependent terms.

The TGV-generated oblique parameters that are produced by this analysis are
displayed in Table VII, where the numerical values $\alpha(\mz^2)=1/128$
and $\sw^2=0.23$ are used. We also take $M=1$ TeV and the effective couplings
are evaluated at the low-energy scale $\mu = 100$ GeV. The quantities $\hat V$
and $\hat X$ incorporate the nonuniversal, $m_t$-dependent, contributions, and
are to be used instead of $V$ and $X$ in the expressions for any observables
which are based on the process $Z \to b\ol{b}$.

The expressions from Table VII can then be used in Table I,  for the
electroweak observables in terms of the oblique parameters. When all five TGV
coefficients are left free in the resulting fit to the data, no useful bound is
obtained, beyond the ever-present ones like perturbative unitarity.
This shows that the data is not yet sufficiently accurate to constrain
these quantities. A common procedure in the literature is to instead bound each
TGV
separately, with all of the others constrained to vanish. This type of analysis
gives, in our case, the following one-$\sigma$ allowed ranges \cite{tgvbounds}:
\eqa
&& \qquad \qquad \dgz = -0.033 \pm 0.031 ; \nn\\
&& \dkg = 0.056 \pm 0.056 ;  \qquad \dkz = -0.0019 \pm 0.044; \\
&& \lg = -0.036 \pm 0.034 ; \qquad \lz = 0.049 \pm 0.045. \nn
\eeqa
Although the bounds obtained in this way are more restrictive, they are also
not realistic for any underlying theory, for which all couplings would be
expected to be generated together.

\section{Conclusions}

This article reviews two approaches to parameterizing the effects of new
physics
for precision electroweak measurements. The two approaches are based on one of
the following two assumptions. Either: ($i$) the new physics is assumed to be
very heavy in comparison with the weak scale, $\mz$, or: ($ii$) it is {\it not}
assumed to be heavy, but it is assumed to dominantly contribute to observables
through oblique corrections. These two approaches contain the popular
formalisms of Peskin \etal, and of Altarelli \etal, as important special cases.

The advantage of parameterizing the data in this way is the efficiency with
which it permits the comparison of specific models to the data. Rather than
having to perform a detailed fit to the data of every proposed theory, it is
possible to fit the data once and for all to the proposed parameters. To
constrain any particular model it is then simply necessary to compute these
parameters in terms of the couplings of the underlying theory. A conservative
estimate of the allowed range for these couplings can be found by simply
using the appropriate confidence intervals for the basic parameterization. The
bounds that are obtained in this way turn out to be remarkably similar to those
which are obtained from model-by-model fits.

A number of phenomenological analyses have been performed using this procedure,
some of which have been briefly summarized here. These calculations illustrate
the
potential applications of the method, and the simplicity with which it may be
carried out.

\section{Acknowledgements}

I would like to thank the organizers of the workshop for their warm
hospitality,
and for their kind invitation to speak. The research described here
was the result of fabulous collaborations with Peter Bamert, Steve Godfrey,
Heinz K\"onig, David London and Ivan Maksymyk, and has been supported by the
Swiss National Foundation, N.S.E.R.C.\ of Canada and les Fonds F.C.A.R.\ du
Qu\'ebec.

\def\pr#1{\it Phys.~Rev.~{\bf #1}}
\def\np#1{\it Nucl.~Phys.~{\bf #1}}
\def\pl#1{\it Phys.~Lett.~{\bf #1}}
\def\prc#1#2#3{{\it Phys.~Rev.~}{\bf C#1} (19#2) #3}
\def\prd#1#2#3{{\it Phys.~Rev.~}{\bf D#1} (19#2) #3}
\def\prl#1#2#3{{\it Phys. Rev. Lett.} {\bf #1} (19#2) #3}
\def\plb#1#2#3{{\it Phys. Lett.} {\bf B#1} (19#2) #3}
\def\npb#1#2#3{{\it Nuc. Phys.} {\bf B#1} (19#2) #3}
\def\zpc#1#2#3{{\it Zeit. Phys.} {\bf C#1} (19#2) #3}
\def\etal{{\it et.al. \/}}

\bibliographystyle{unsrt}

\begin{thebibliography}{99}
\bibitem{top}
At this writing some evidence for the existence of a $t$ quark having a mass of
roughly 170 GeV has been reported by: F. Abe \etal, \prl{73}{94}{225};
\prd{50}{94}{2966}.
\end{thebibliography}

\pagebreak

\begin{center}
\begin{tabular}{ll}
%\noalign{\medskip}\noalign{\hrule}\noalign{\smallskip}
%\noalign{\hrule}\noalign{\medskip}
$\Gamma_\ssz  =(\Gamma_\ssz)_{\sss SM} - 0.00961 S + 0.0263 T
 + 0.0194 V - 0.0207 X $  (GeV)  & \\
$\Gamma_{b \ol{b}}  =(\Gamma_{b\ol{b}})_{\sss SM} - 0.00171 S + 0.00416 T
 + 0.00295 V - 0.00369 X$ (GeV)  & \\
$\Gamma_{l^+ l^-}=(\Gamma_{l^+ l^-})_{SM} -0.000192 S
  + 0.000790 T + 0.000653 V - 0.000416 X $ (GeV)  &\\
$\Gamma_{had}=(\Gamma_{had})_{SM} -0.00901 S
  + 0.0200 T + 0.0136 V - 0.0195 X $ (GeV) &\\
$A_{\sss FB}(\mu)= (A_{\sss FB}(\mu))_{\sss SM} - 0.00677 S +
0.00479 T - 0.0146 X$  & \\
$A_{pol}(\tau) = (A_{pol}(\tau))_{\sss SM} -0.0284 S + 0.0201 T - 0.0613
X $  & \\
$A_e (P_\tau) =(A_e(P_\tau))_{\sss SM} -0.0284 S + 0.0201 T - 0.0613 X $ &\\
$A_{\sss FB}(b)=(A_{\sss FB}(b))_{\sss SM} -0.0188 S + 0.0131 T
-0.0406X$  & \\
$A_{\sss FB}(c)=(A_{\sss FB}(c))_{\sss SM} -0.0147 S + 0.0104 T -0.03175
X$  &\\
$A_{\sss LR} =(A_{\sss LR})_{\sss SM} -0.0284 S + 0.0201 T - 0.0613 X $ & \\
$\mw^2=(\mw^2)_\sm (1-0.00723 S +0.0111 T +0.00849 U)$ &\\
$\Gamma_\ssw =(\Gamma_\ssw)_{\sss SM}(1-0.00723 S + 0.0111 T + 0.00849 U
+ 0.00781W) $ & \\  %% this fixes a sign (for the T term) from the vwxfit
% paper.
$g_{\sss L}^2 =(g_{\sss L}^2)_{\sss SM}-0.00269 S + 0.00663 T$ & \\
$g_{\sss R}^2 =(g_{\sss R}^2)_{\sss SM} +0.000937 S - 0.000192  T$ & \\
$g_{\sss V}^e(\nu e \to \nu e) =(g_{\sss V}^e)_{\sss SM} + 0.00723 S -
0.00541 T$  & \\
$g_{\sss A}^e (\nu e \to \nu e) =(g_{\sss A}^e)_{\sss SM} - 0.00395 T$ & \\
$Q_\ssw(^{133}_{55}Cs) = Q_\ssw(Cs)_{\sss SM} -0.795 S -0.0116 T$ & \\
%\noalign{\medskip}\noalign{\hrule}\noalign{\smallskip}
%\noalign{\hrule}\noalign{\medskip}
\end{tabular}
\end{center}
\begin{center} {\bf TABLE I: Oblique Contributions to Observables} \end{center}
\medskip\noindent
The dependence of some electroweak observables on $S,T,U,V,W$ and $X$.
The numerical values $\alpha(\mz^2)=1/128$ and $\sw^2=0.23$ are used in
preparing this table. The precise definitions of the observables can be found
in Refs.~\citenum{alphabet}.

\pagebreak

\begin{center}
\begin{tabular}{l}
$R_\pi \equiv \Gamma(\pi\to e\nu) / \Gamma(\pi\to\mu\nu)
= R_\pi^\sm \left( 1 + 2 \Delta_e - 2 \Delta_\mu \right)$ \\
$R_{\tau} \equiv \Gamma(\tau\to e\nu\bar{\nu})   /
\Gamma(\mu\to e\nu{\overline\nu})  = R_\tau^\sm
\left(1 + 2 \Delta_\tau - 2 \Delta_\mu \right) $\\
$R_{\mu\tau} \equiv \Gamma(\tau\to\mu\nu{\overline\nu}) /
\Gamma(\mu\to e\nu{\overline\nu})  = R_{\mu\tau}^\sm \left(1 + 2
\Delta_\tau - 2 \Delta_e \right) $\\
$\rho = 1 + \alpha T $\\
$ \sigma(\nu N\to\mu^- X) = \sigma^\sm(\nu N\to\mu^- X)  \Bigl[
1 + 2\Delta_\mu - 2\Delta_e - 2\;  (\Re(\delta \twh_\rht^{ud}) /
  \vert V_{ud} \vert) \Bigr] $\\
$\left(g_\lft^2 \right) = \left( g^2_\lft \right)_\sm - 0.00269 \, S
 + 0.00663 \, T -1.452 \Delta_\mu  - 0.244 \Delta_e $ \\
\qquad\qquad $+ 0.620 \, \Re(\delta \twh_\rht^{ud})
- 0.856 \, \delta \twg_\lft^{dd} + 0.689 \, \delta \twg_\lft^{uu}
+ 1.208 \, \delta \twg_\lft^{\nu_\mu\nu_\mu} $\\
$\left(g_\rht^2\right) = \left(g_\rht^2\right)_\sm
+ 0.000937 \, S - 0.000192 \, T + 0.085 \Delta_e - 0.0359 \Delta_\mu $\\
\qquad\qquad $+ 0.0620 \, \Re(\delta \twh_\rht^{ud})
+ 0.156 \, \delta \twg_\rht^{dd} - 0.311 \, \delta \twg_\rht^{uu}
+ 0.121 \, \delta \twg_\lft^{\nu_\mu\nu_\mu}$\\
$g_{e\ssv} = \left( g_{e\ssv}\right)_\sm + 0.00723 \, S  - 0.00541 \, T +
0.656 \Delta_e + 0.730 \Delta_\mu $\\
\qquad\qquad $+ \delta \twg_\lft^{ee} + \delta \twg_\rht^{ee}  - 0.074 \,
\delta
  \twg_\lft^{\nu_\mu\nu_\mu} - 0.037 \, \Re(\delta\twh_\rht^{ud}) $\\
$g_{e\ssa} = \left( g_{e\ssa} \right)_\sm - 0.00395 \, T + 1.012 \Delta_\mu
  + \delta \twg_\lft^{ee} - \delta \twg_\rht^{ee} - 1.012 \, \delta
  \twg_\lft^{\nu_\mu\nu_\mu} - 0.0506 \, \Re(\delta\twh_\rht^{ud}) $\\
$C_{1u} = C_{1u}^\sm + 0.00482 \, S - 0.00493 \, T
+ 0.631 (\Delta_e + \Delta_\mu) $\\
\qquad\qquad $+ 0.387 \,\delta \twg_\lft^{ee} - \delta \twg_\lft^{uu}
- 0.387 \,\delta \twg_\rht^{ee} - \delta \twg_\rht^{uu}$ \\
$C_{1d} = C_{1d}^\sm - 0.00241 \, S +  0.00442 \, T
- 0.565 (\Delta_e + \Delta_\mu) $\\
\qquad\qquad $- 0.693 \,\delta \twg_\lft^{ee} - \delta \twg_\lft^{dd}
+ 0.693 \,\delta \twg_\rht^{ee} - \delta \twg_\rht^{dd} $\\
$C_{2u} = C_{2u}^\sm + 0.00723 \, S - 0.00544 \, T
+ 0.696 (\Delta_e + \Delta_\mu) $\\
\qquad\qquad $+ \delta \twg_\lft^{ee} - 0.08 \,\delta \twg_\lft^{uu}
+ \delta \twg_\rht^{ee} + 0.08 \,\delta \twg_\rht^{uu}$ \\
$C_{2d} = C_{2d}^\sm - 0.00723 \, S + 0.00544 \, T
- 0.696 (\Delta_e + \Delta_\mu) $\\
\qquad\qquad $- \delta \twg_\lft^{ee} - 0.08 \,\delta \twg_\lft^{dd}
- \delta \twg_\rht^{ee} + 0.08 \,\delta \twg_\rht^{dd} $\\
$\qw({}^{133}_{55}Cs) = \left[ \qw({}^{133}_{55}Cs) \right]_\sm
- 0.796 \, S -0.0113 \, T + 1.45 (\Delta_e + \Delta_\mu)
+ 147 \left( \delta \twg_\lft^{ee} - \delta \twg_\rht^{ee}\right) $\\
\qquad\qquad $+ 422 \left( \delta \twg_\lft^{dd} + \delta \twg_\rht^{dd}\right)
+ 376 \left( \delta \twg_\lft^{uu} + \delta \twg_\rht^{uu}\right) $\\
\end{tabular}
\end{center}
\begin{center} {\bf Table II: Low-Energy Observables} \end{center}
\medskip\noindent
The contributions to low-energy ($q^2 \approx 0$) electroweak observables
that are generated by the various interactions of the general non-oblique
effective lagrangian.
The precise definitions of the observables can be found in
Ref.~\citenum{bigfit}.

\pagebreak

\begin{center}
\begin{tabular}{l}
$\mw^2 = (\mw^2)_\sm [1 - 0.00723 \, S + 0.0111 \, T +0.00849 \, U
-0.426(\Delta_e+\Delta_\mu)]$ \\
$\Gamma_{\ell^+\ell^-} = \left( \Gamma_{\ell^+ \ell^-}\right)_\sm
\Bigl[ 1 - 0.00230 \, S + 0.00944 T - 1.209 (\Delta_e + \Delta_\mu) $\\
\qquad\qquad $-4.29 \,\delta \twg_\lft^{\ell\ell} + 3.66 \,\delta
\twg_\rht^{\ell\ell} \Bigr] $ \\
$\Gamma_{u{\bar u}} = (\Gamma_{u{\bar u}})_\sm  \Bigl[ 1 - 0.00649 \, S +
0.0124 T - 1.59  (\Delta_e+\Delta_\mu) $\\
 \qquad\qquad $+ 4.82 \,\delta
\twg_\lft^{uu} - 2.13 \,\delta \twg_\rht^{uu} \Bigr] $\\
$\Gamma_{d{\bar d}} = (\Gamma_{d{\bar d}} )_\sm \Bigl[ 1 - 0.00452\, S +
  0.0110 T - 1.41 (\Delta_e + \Delta_\mu) $\\
\qquad\qquad $- 4.57 \,\delta \twg_\lft^{dd} + 0.828 \,\delta \twg_\rht^{dd}
\Bigr]$ \\
$\Gamma_{b{\bar b}} = (\Gamma_{b{\bar b}})_\sm \Bigl[ 1 - 0.00452\, S +
  0.0110 T - 1.41 (\Delta_e + \Delta_\mu) $\\
\qquad\qquad $- 4.57 \,\delta \twg_\lft^{bb} + 0.828 \,\delta \twg_\rht^{bb}
\Bigr] $\\
$\Gamma_{\rm had} = (\Gamma_{\rm had})_\sm
\Bigl[ 1- 0.00518 \, S + 0.0114 T -
1.469 (\Delta_e + \Delta_\mu) $\\
\qquad\qquad $- 1.01 \Bigl( \delta \twg_\lft^{dd} + \delta \twg_\lft^{ss} +
\delta \twg_\lft^{bb} \Bigr) + 0.183 \Bigl( \delta \twg_\rht^{dd} + \delta
  \twg_\rht^{ss} + \delta \twg_\rht^{bb} \Bigr)  $\\
\qquad\qquad $+ 0.822 \Bigl( \delta \twg_\lft^{uu} + \delta \twg_\lft^{cc}
\Bigr)
- 0.363 \Bigl( \delta \twg_\rht^{uu} + \delta \twg_\rht^{cc} \Bigr)
\Bigr] $\\
$\Gamma_{\nu_i{\overline\nu}_i}
= (\Gamma_{\nu_i{\overline\nu}_i})_\sm \Bigl[ 1 +0.00781 T -
(\Delta_e+\Delta_\mu) + 4 \,\delta \twg_\lft^{\nu_i \nu_i} \Bigr] $ \\
$\Gamma_{\sss Z} = (\Gamma_\ssz)_\sm \Bigl[ 1 - 0.00385 \, S +0.0105 \, T
  - 1.35 (\Delta_e+\Delta_\mu) + 0.574 \left( \delta \twg_\lft^{uu} +
  \delta \twg_\lft^{cc} \right) $\\
\qquad\qquad $- 0.254 \left( \delta \twg_\rht^{uu} + \delta \twg_\rht^{cc}
 \right)  + 0.268 \left( \delta \twg_\lft^{\nu_e\nu_e} + \delta
 \twg_\lft^{\nu_\mu\nu_\mu} + \delta \twg_\lft^{\nu_\tau\nu_\tau}
 \right) $\\
\qquad\qquad $ - 0.144 \left( \delta \twg_\lft^{ee} + \delta \twg_\lft^{\mu\mu}
+
  \delta \twg_\lft^{\tau\tau} \right) + 0.123 \left( \delta \twg_\rht^{ee}
  + \delta \twg_\rht^{\mu\mu} + \delta \twg_\rht^{\tau\tau}
\right) $\\
\qquad\qquad $- 0.707 \left( \delta \twg_\lft^{dd} + \delta \twg_\lft^{ss} +
  \delta \twg_\lft^{bb} \right) + 0.128 \left( \delta \twg_\rht^{dd} +
  \delta \twg_\rht^{ss} + \delta \twg_\rht^{bb} \right) \Bigr] $\\
$A_{\lft\rht}  = A_{\lft\rht}^\sm  - 0.0284 \, S + 0.0201 \, T - 2.574 (
\Delta_e +
  \Delta_\mu) - 3.61 \, \delta \twg_\lft^{ee} - 4.238 \, \delta
\twg_\rht^{ee}$\\
$A_{\sss FB}^{\ell^+\ell^-} =  (A_{\sss FB})_\sm - 0.00677 \, S
+ 0.00480 \, T - 0.614  (\Delta_e + \Delta_\mu) $\\
\qquad\qquad $-0.430 \left( \delta \twg_\lft^{ee} + \delta
\twg_\lft^{\ell\ell} \right) -0.505 \left( \delta \twg_\rht^{ee} + \delta
\twg_\rht^{\ell\ell} \right) $\\
$A_{\sss FB}(b{\bar b}) =  (A_{\sss FB}(b{\bar b}))_\sm -0.0188 \, S
+ 0.0133 \, T - 1.70 (\Delta_e  +\Delta_\mu) $\\
 \qquad\qquad $-2.36 \,\delta \twg_\lft^{ee} - 2.77 \,\delta \twg_\rht^{ee}
  -0.0322 \,\delta \twg_\lft^{bb} - 0.178 \,\delta \twg_\rht^{bb} $\\
$A_{\sss FB}(c{\bar c}) = (A_{\sss FB}(c{\bar c}))_\sm - 0.0147 \, S
 + 0.0104 \, T - 1.333( \Delta_e + \Delta_\mu) $\\
\qquad\qquad $- 1.69 \,\delta \twg_\lft^{ee} - 1.99 \,\delta \twg_\rht^{ee}
   + 0.175 \,\delta \twg_\lft^{cc}  + 0.396 \,\delta \twg_\rht^{cc} $\\
\end{tabular}
\end{center}
\begin{center} {\bf Table III: Weak-Scale Observables} \end{center}
\medskip\noindent
The contributions to weak-scale ($q^2 = \mz^2$ or $\mw^2$) electroweak
observables
that are generated by the various interactions of the general non-oblique
effective lagrangian. The precise definitions of the observables can be
found in Ref.~\citenum{bigfit}.

\pagebreak

\begin{center}
\begin{tabular}{ccc}
Parameter & Individual Fit & Global Fit \\
&&\\
$S$ & $-0.10 \pm 0.16$ & $-0.2 \pm 1.0$ \\
$T$ & $+0.01 \pm 0.17$ & $-0.02 \pm 0.89$ \\
$U$ & $-0.14 \pm 0.63$ & $+ 0.3 \pm 1.2$ \\
\end{tabular}
\end{center}
\begin{center} {\bf Table IV: Oblique Parameters} \end{center}
\medskip\noindent
Results for the oblique parameters $S$, $T$ and $U$ obtained from the
fit of the new-physics parameters to the data. The second column gives
the result for the (unrealistic) case where all other parameters are
constrained
to vanish. Column three gives the result of a global fit in which all of the
parameters of the effective lagrangian are varied.

\pagebreak

\begin{center}
\begin{tabular}{ccc}
Parameter & Individual Fit & Global Fit \\
&&\\
$\Delta_e$ & $-0.0008\pm .0010$  & $-0.0011 \pm .0041$ \\
$\Delta_\mu$ & $+0.00047 \pm .00056$  & $+0.0005\pm .0039$ \\
$\Delta_\tau$ & $-0.018 \pm 0.008$ & $-0.018\pm .009$ \\
$\Re(\delta\twh^{ud}_\lft)$ & $-0.00041\pm .00072$ & $+0.0001\pm
.0060$\\
$\Re(\delta\twh^{ud}_\rht)$ & $-0.00055\pm .00066$ & $+0.0003 \pm
.0073$\\
$\Im(\delta\twh^{ud}_\rht)$ & $0 \pm 0.0036$ & $-0.0036 \pm .0080$\\
$\Re(\delta\twh^{us}_\lft)$ & $-0.0018\pm .0032$ & --- \\
$\Re(\delta\twh^{us}_\rht)$ & $-0.00088\pm .00079$ & $+0.0007\pm
.0016$\\
$\Im(\delta\twh^{us}_\rht)$ & $0 \pm 0.0008$ & $-0.0004\pm .0016$\\
$\Re(\delta\twh^{ub}_\lft)$,$\Im(\delta\twh^{ub}_\lft)$
& $-0.09\pm .16$ & --- \\
$\Sigma_1$ & --- &$+0.005\pm .027$ \\
$\Re(\delta\twh^{ub}_\rht)$ & --- & --- \\
$\Re(\delta\twh^{cd}_\lft)$ & $+0.11 \pm .98$ & --- \\
$\Re(\delta\twh^{cd}_\rht)$ & --- & --- \\
$\Re(\delta\twh^{cs}_\lft)$ & $+0.022\pm .20$ & --- \\
$\Re(\delta\twh^{cs}_\rht)$ & $+0.022\pm .20$ & --- \\
$\Re(\delta\twh^{cb}_\lft)$ & $+0.5\pm 4.6$ & --- \\
$\Sigma_2$ & --- & $+0.11\pm 0.98$ \\
$\Re(\delta\twh^{cb}_\rht)$ & --- & ---  \\
\end{tabular}
\end{center}
\begin{center} {\bf Table V: Charged-Current Parameters} \end{center}
\medskip\noindent
More results of the fits of the
new-physics parameters to the data. The quantities $\Sigma_1$ and
$\Sigma_2$ arise in tests for the unitarity of the CKM matrix,
and are defined as: $\Sigma_1 \equiv \Re( \delta \twh_\lft^{us}) +
\left[\Re(V_{ub}) \Re(\delta \twh_\lft^{ub}) + \Im(V_{ub}) \Im(\delta
\twh_\lft^{ub}) \right]/ \vert V_{us}\vert $ and $\Sigma_2 \equiv
\Re(\delta \twh_\lft^{cd}) + \vert V_{cs} \vert \Re(\delta \twh_\lft^{cs} +
\delta \twh_\rht^{cs})/\vert V_{cd} \vert + |V_{cb}| \Re(\delta
\twh_\lft^{cb})/\vert V_{cd} \vert $. Blanks indicate where the
corresponding fit would be inappropriate, such as for when a parameter
always appears in a particular combination with others, and so
cannot be individually fit.

\pagebreak

\begin{center}
\begin{tabular}{ccc}
Parameter & Individual Fit & Global Fit \\
&&\\
$\delta\twg^{dd}_\lft$ & $+0.0016\pm .0015$ & $+0.003\pm .012$ \\
$\delta\twg^{dd}_\rht$ & $+0.0037\pm .0038$ & $+0.007\pm .015$ \\
$\delta\twg^{uu}_\lft$ & $-0.0003\pm .0018$ & $-0.002\pm 0.014$ \\
$\delta\twg^{uu}_\rht$ & $+0.0032\pm .0032$ & $-0.003 \pm .010$ \\
$\delta\twg^{ss}_\lft$ & $-0.0009\pm .0017$ & $-0.003\pm .015$ \\
$\delta\twg^{ss}_\rht$ & $-0.0052\pm .00095$ & $+0.002\pm .085$ \\
$\delta\twg^{cc}_\lft$ & $-0.0011\pm .0021$ & $+0.001\pm .018$ \\
$\delta\twg^{cc}_\rht$ & $+0.0028\pm .0047$ & $+0.009\pm .029$ \\
$\delta\twg^{bb}_\lft$ & $-0.0005\pm .0016$ & $-0.0015\pm .0094$ \\
$\delta\twg^{bb}_\rht$ & $+0.0019\pm .0083$ & $0.013\pm .054$ \\
 $\delta\twg^{\nu_e \nu_e}_\lft$ & $-0.0048\pm .0052$ & --- \\
$\delta\twg^{\nu_\mu \nu_\mu}_\lft$ & $-0.0021\pm .0027$ &
	$+0.0023 \pm .0097$ \\
$\delta\twg^{\nu_\tau \nu_\tau}_\lft$ & $-0.0048\pm .0052$ &--- \\
$\delta\twg^{\nu_e \nu_e}_\lft + \delta\twg^{\nu_\tau \nu_\tau}_\lft$ &
	--- & $-0.004 \pm .033$ \\
$\delta\twg^{ee}_\lft$ & $-0.00029\pm .00043$ & $-0.0001\pm .0032$ \\
$\delta\twg^{ee}_\rht$ & $-0.00014\pm .00050$ & $+0.0001\pm .0030$ \\
$\delta\twg^{\mu\mu}_\lft$ & $+0.0040\pm .0051$ & $+0.005\pm .032$ \\
$\delta\twg^{\mu\mu}_\rht$ & $-0.0003\pm .0047$ & $+0.001\pm .028$ \\
$\delta\twg^{\tau\tau}_\lft$ & $-0.0021\pm .0032$ & $\; 0.000\pm .022$
\\
$\delta\twg^{\tau\tau}_\rht$ & $-0.0034\pm .0028$ &
	$-0.0015\pm .019$ \\
\end{tabular}
\end{center}
\begin{center} {\bf Table VI: Neutral-Current Parameters} \end{center}
\medskip\noindent
Still more results of the fits of the new-physics parameters to the data.
As before, blanks indicate where the
corresponding fit would be inappropriate, such as for when a parameter
always appears in a particular combination with others, and so
cannot be individually fit.

\pagebreak

\begin{center}
\begin{tabular}{cc}
Parameter & One-Loop Result \\
&\\
$S$ & $2.63 \dgz - 2.98 \dkg + 2.38 \dkz + 5.97 \lg - 4.50\lz $\\
$T$ & $-1.82 \dgz + 0.550 \dkg + 5.83 \dkz $\\
$U$ & $ 2.42 \dgz - 0.908 \dkg - 1.91 \dkz + 2.04 \lg - 2.04 \lz $\\
$V$ & $ 0.183 \dkz  $\\
$W$ & $ 0.202 \dgz  $\\
$X$ & $ -0.0213 \dkg - 0.0611 \dkz $\\
$\hat{V}$ & $ 0.183 \dkz - (3.68 \dgz + 0.797 \dkz) (m_t^2/\mw^2)  $\\
$\hat{X}$ & $ -0.0213 \dkg - 0.0611 \dkz +
(0.423 \dgz + 0.0916 \dkz) (m_t^2/\mw^2)  $\\
\end{tabular}
\end{center}
\begin{center} {\bf TABLE VII: TGV Contributions to Oblique Parameters}
\end{center}

\bigskip\noindent
One-loop results for the induced parameters $S$, $T$,
$U$, $V$, $W$ and $X$, defined at $\mu = 100$ GeV, in terms of the various
TGV couplings defined at $M=1$ TeV.  As usual, $\alpha(\mz^2)=1/128$ and
$\sw^2=0.23$.

\pagebreak

\begin{center} {\bf Figure Captions} \end{center}

\begin{enumerate}
\item[{\bf Figure 1:}]
Constaints on $S$ and $T$ from a fit to both high- and low-energy
electroweak measurements. The solid line represents the 68\% C.L. setting
$VWX$ to zero, the dashed line represents the 90\% C.L. setting $VWX$ to
zero, the dotted line represents the 68\% C.L. allowing $VWX$ to vary, and
the dot-dashed line represents the 90\% C.L. allowing $VWX$ to vary.
\item[{\bf Figure 2:}]
Constaints on $S'$ and $T'$ from a global fit of precison
electroweak measurements at the $Z$ resonance only. The two
lines represent the 68\% C.L. and 90\% C.L. ellipses.
\item[{\bf Figure 3:}]
The contribution of an extra SM family to the oblique parameters $S'$
and $T'$. Solid line: a colour-triplet, $Y=\nth{6}$ quark doublet with masses
$(m_{b'},m_{t'})$ as indicated in brackets. Dotted line: a colour-singlet,
$Y={1 \over 2}$ lepton doublet with masses $(m_{\nu '},m_{l'})$. The grid
spacing
represents steps of 25 (resp. 30) GeV for the solid (resp. dotted)  plots.
\end{enumerate}

\end{document}